\documentclass[%
 reprint,
 amsmath,amssymb,
 aps,
pra,
]{revtex4-1}

\usepackage{grffile}
\usepackage{xcolor}
\usepackage{epstopdf}
\usepackage{graphicx}% Include figure files
\usepackage{dcolumn}% Align table columns on decimal point
\usepackage{subfigure}
\usepackage{bm}% bold math
\def\rev{black}
\def\revb{\color{black}}
\usepackage[normalem]{ulem}

\def\revr{\color{black}}

\def\revrr{\color{black}}

\begin{document}

%\preprint{APS/123-QED}

\title{\textcolor{\rev}{Modal analysis of  wave propagation in dispersive media}}

% Force line breaks with \\

\author{M. Ismail Abdelrahman}
%\author{Iva}%
 \email{m.abdelrahman@fresnel.fr}
\affiliation{%
Aix Marseille Univ, CNRS, Centrale Marseille, Institut Fresnel, Marseille, France}%
\author{B. Gralak}
%\author{Iva}%
% \email{m.abdelrahman@fresnel.fr}
\affiliation{%
  Aix Marseille Univ, CNRS, Centrale Marseille, Institut Fresnel, Marseille, France}%

\begin{abstract}
Surveys on wave propagation in dispersive media have been limited since the pioneering work of
Sommerfeld [Ann. Phys. {\textbf{349}}, 177 (1914)] by the presence of branches in the integral expression
of the wave function. {\revr{In this {\revr{article}}, a method is proposed to eliminate these critical branches and hence to establish a
{\revb{modal expansion}} of the time-dependent wave function. }} The different components of the 
transient waves are physically interpreted as the contributions of distinct sets of {\revb{modes}} and   
characterized accordingly. {\revb{Then, the modal expansion is used to derive a modified analytical expression 
of the Sommerfeld precursor improving significantly  the description of the amplitude and the
oscillating period up to the arrival of the Brillouin precursor.}}
The proposed method and results apply to all waves governed by the Helmholtz equations.

\end{abstract}

%\pacs{\pacs{meta material}}% PACS, the Physics and Astronomy
%\keywords{gg keywords}%Use showkeys class option if keyword
\maketitle

%\tableofcontents

%------------------------------------------------------------------------------------------------
% notations here:
% notations here:
\def\ep{\varepsilon}
\def\eps{\varepsilon_s}
\def\om{\omega}
\def\oms{\om_s}
\def\omr{\om_0}
\def\omp{\om_p}
\def\omq{\om_q}
\def\Om{\Omega}
\def\hom{\hat\omega}
\def\homs{\hat\om_s}
\def\homr{\hat\om_0}
\def\homp{\hat\om_p}
\def\homq{\hat\om_q}
\def\hOm{\hat\Omega}
\def\r{r}
\def\xs{x_0}
\def\phase{\phi}
\def\R{R}
\def\T{T}

\def\b{\color{blue}}
\def\br{\color{red}}
%----------------------

\section{\label{sec1}Introduction}
The modern developments of wave propagation in linear homogeneous dispersive media have been initiated by Sommerfeld \cite{sommerfeld1914fortpflanzung} and Brillouin \cite{brillouin1960wave}. The main motivation was to investigate the causality principle \cite{stenner2003speed}, when strong dispersion holds, and the notion of signal velocity \cite{brillouin1960wave}. These pioneering works led to the description of the Sommerfeld  and Brillouin precursors (also known as forerunners), which are transient waves preceding the main propagating signal.
Analyzing these precursors is unavoidable for a wideband signal propagating in dispersive media since, for instance, they are  responsible for the pulse distortion in communication systems  \cite{oughstun1990uniform, cartwright2007uniform}.  Moreover, they can be  useful for  probing biological samples \cite{albanese1989short},  for underwater communications \cite{choi2004observation}, and for warning systems based on the early detection of precursors \cite{renzi2014hydro}. 
 They have been  observed  at the microwave and optical frequencies \cite{pleshko1969experimental,aaviksoo1991observation,sakai2002polariton,choi2004observation,jeong2006direct}.
%, in slow light \cite{du2008observation}, and even in  the level of a single photon \cite{zhang2011optical}. 
Similar behavior occurs in other situations such as spatially-dispersive media \cite{frankel1977transient}, 
photonic crystals \cite{uitham2006sommerfeld}, acoustic waves 
%and on the surface of fluids \cite{zhu2014acoustic,falcon2003observation,varoquaux1986pulse}
\cite{varoquaux1986pulse,falcon2003observation}, gravity waves \cite{renzi2014hydro}, or more generally, all waves  described by Helmholtz equations.  More recently, the introduction of negative index materials \cite{Ves68,Pendry00,Notomi00,Gralak00}, metamaterials and transformation optics \cite{pendry2006controlling,Leonhardt06,schurig2006metamaterial}  caused an additional interest for the dispersion phenomenon \cite{Ves68}. Hence, the effect of dispersion has been recently investigated in the cases of the flat lens  \cite{GT10,Collin10,PRL-Pen11,GM12,PRL-Gref12} and invisibility systems \cite{Gra16}.

Since 1914 and the work of Sommerfeld \cite{sommerfeld1914fortpflanzung}, the analysis of wave propagation in dispersive media has been restricted by the presence of branch-cuts in \textcolor{\rev}{ the solution to the wave equation for the electric field $E$} \cite{oughstun1997failure}. Let a sinusoidal source of frequency $\om_s$ be switched on at an initial time $t = 0$ in {\revr {an infinite}} dispersive medium of relative permittivity $\ep(\om)$. Then, the temporal behavior of the field at a distance $x$ from the source is given by the following integral in the complex frequency domain \cite{brillouin1960wave}:
\begin{equation}  \label{eq:1maineq}
{\revb{E_\infty}}(x,t)= {\revb{-}} \frac{E_0 }{2 \pi}\displaystyle\int_\Gamma d\om\,  
e^{- i \om t} \, \frac{\oms}{\om^2 - \oms^2} \; \; e^{i \om \sqrt{\ep(\om)} x / c },
\end{equation}
where $E_0$ is proportional to the source amplitude at $x = 0$, $c$ is the light velocity in vacuum, $\sqrt{\ep(\om)}$ is the refractive index of the medium, and the integration path $\Gamma$ is a line parallel to the real axis in the half plane of complex frequencies with positive imaginary part. The main difficulty to compute the integral (1) is the presence of the square root $\sqrt{\ep(\om)}$, which induces branch-points and branch-cuts constituting a severe drawback for a direct evaluation.
This issue has been mainly addressed using numerical methods \cite{alejos2012temporal} and asymptotic approaches \cite{brillouin1960wave,handelsman1969uniform,oughstun1988propagation,Oug89,MS12},  the latter requiring a specific form of the dispersion relation \cite{oughstun1991pulse,macke2009optical}. As for the group velocity approximation, it usually fails to analyze wideband pulses \cite{oughstun1997failure}. The book of Oughstun \cite{Oughstun2009} can be consulted for an extensive bibliography.

%------------------------------------------------------------------------------------------------
%  Fig 1
%------------------------------------------------------------------------------------------------
\begin{figure}[b!]
\includegraphics[ width=0.95\linewidth, keepaspectratio]{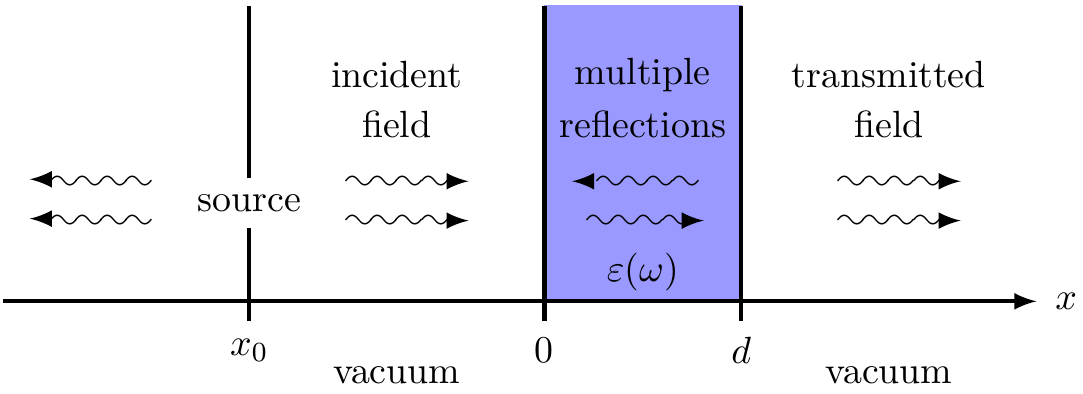}
\caption{A  slab of a homogeneous  medium of arbitrarily dispersion  is illuminated in normal incidence (1-D problem). } \label{Fig1}
\end{figure}
%------------------------------------------------------------------------------------------------

\textcolor{\rev}{In this {\revr{article}}, a novel approach is proposed to remove this critical branch-cut{\revr{:}}
instead of wave propagation in an infinite {\revr{homogeneous}} medium, wave transmission through a slab of a thickness 
$d$ {\revb{is considered}} (see Fig. \ref{Fig1}). In this case, {\revb{it is shown}} that no {\revb{branches}} 
exist in the integral expression {\revr{  and, consequently,    the transmitted field can be evaluated in the time domain using a discrete modal expansion}}. This {\revr{modal expansion}} is then 
used to interpret and characterize all the components of the transient waves, e.g. the Sommerfeld and 
Brillouin precursors.  {\revr{ Interestingly, the transient waves  at the onset of the transmitted field with velocities close to $c$,  which experience a near unity permittivity, are matching their counterparts  in infinite media since the slab reflections  are negligible. {\revr{This striking resemblance}} allows the use of  the proposed modal expansion to  derive  a  modified analytical expression of the Sommerfeld precursor which,  compared to the original expression \cite{sommerfeld1914fortpflanzung}, improves significantly the amplitude and periodicity description.}} }

\section{\label{sec2}modal expansion}

The considered system is made of a dispersive slab with permittivity $\ep(\om)$ and thickness $d$, as represented in Fig. \ref{Fig1}. Normalized \textcolor{\rev}{(dimensionless)} quantities are used to describe the problem in a simple way, 
\begin{equation} 
\ \om \; (d  / c) \longrightarrow \om \, , \quad
 (t - |x_0|/c) \; (c / d)  \longrightarrow  t \, .
\label{hom}
\end{equation}
%\cite{premaratne2011light}. A
Assuming the surrounding medium to be vacuum, the system can be treated as a symmetric passive Fabry-P\'erot resonator with the multiple reflections effect. The reflection coefficient $\rho$ of the slab interface is 
\begin{equation}\label{eq:r}
\rho(\om) = \frac{1 - \sqrt{\ep(\om)} } {1 + \sqrt{\ep(\om)} } \, ,
\end{equation}
and the transmitted field can be analyzed using the resonator transmission 
function $T(\om)$ \cite{cesini1977response}:
\begin{equation} \label{eqSlabTF}
T(\om) = \frac{[ 1 - \rho^2(\om) ] \, e^{i \om \sqrt{\ep(\om)} }}{1-\rho^2(\om) \, 
e^{2 i \om \sqrt{\ep(\om)}}}.
\end{equation}
It is stressed that, after some algebra, this transmission function can be expressed as
%\begin{equation}\label{eqT}
%T^{-1}(\om) = \cos[\om \sqrt{\ep(\om)}] - %0.5\; i \, [1 + \ep(\om)] \, 
%\sin[\om \sqrt{\ep(\om)}] \; /\sqrt{\ep(\om)},
%\end{equation}
\begin{equation}\label{eqT}
\frac{1}{T(\om)} = \cos[\om \sqrt{\ep(\om)}] - i \, \frac{1 + \ep(\om)}{2} \, 
\frac{\sin[\om \sqrt{\ep(\om)}]}{\sqrt{\ep(\om)}},
\end{equation}
excluding the case $T=0$, which is in practice irrelevant to the study herein. Indeed, this expression shows that $T(\om)$ is an even function of $\sqrt{\ep(\om)}$; this important remark implies the absence of branches of square roots of the permittivity in $T(\om)$. 
Similar to Eq. (\ref{eq:1maineq}), the time dependence of the transmitted field at the output is 
\begin{equation} 
E_T(t) = - \frac{E_0}{2 \pi}\displaystyle\int_\Gamma d\om\,  \, 
e^{- i \om t} \, \frac{\oms}{\om^2 - \oms^2} \, T(\om) \, ,
\label{intslab}
\end{equation}
where $ E_0(\om)$ is the incident field  at $x=0$. This integration contains neither branch-points nor  branch-cuts, which is the key 
argument of this study. \textcolor{\rev}{Accordingly, it is possible to derive a closed-form expression 
of the transmitted field  in terms of the contributions of the poles of the integrand, {\revr{corresponding to a modal expansion}}}.  %That allows a straightforward physical interpretation of different components of the transmitted fields since each pole represents  a mode of this system. That is a crucial advantage of the {\revb{modal approach}} in comparison with the method usually used for dispersive media \cite{brillouin1960wave}.

\textcolor{\rev}{The transmitted field {\revr{is}} then resolved into two main contributions: 
$E_s(t)$ the contribution of the source poles $\pm\omega_s$, and $E_r(t)$ the contribution 
of {\revb{the}} resonator {\revb{poles}} $\{\om_q, q \text{ integer}\}$ of  $T(\omega)$. Namely, }

\begin{equation} \label{twoparts} 
{\color{\rev}
E_T(t) =  \theta(t -  t_0)\, E_{\textrm{s}}(t) +  \theta(t - t_0)\, 
E_{\textrm{r}}(t),}
\end{equation}
where $\theta$ is Heaviside unit step function; $ \theta(t) =0$ for $ t < 0$ and 
$ \theta(t) =1$ for $ t > 0$.  In normalized units, $t_0 = 1$ is the time needed for 
the front of the wave %(the high-frequency components of the source spectrum that experience a unity refractive index, thus a group velocity $c_o$ \cite{brillouin1960wave}) 
to reach from the input to the output of the slab. That preserves causality since no 
signal can travel faster than light.
% If we assume a lossless source where the poles are located on the real axis of the complex frequency domain,  $E_s(t)$  constitutes the steady-state solution  $E_T(t \to \infty)$.  The contribution of the resonator poles $E_r(t)$ are responsible for the transient behavior, where the decay rate each of pole term depends on the imaginary part of the pole, as  shown  in Eq. (\ref{eq:Transient}).  }
Using the Hermitian property of the 
transmission function $\overline{T(\omega)} = T(-\overline{\omega})$, one  obtains
\begin{equation} \label{eq:stst} 
\vspace*{2mm}
E_{\textrm{s}}(t)= - \text{Im} 
\big[ e^{- i \oms t }\; T(\oms) E_0(\oms) \, \big] \, , 
\end{equation}
where ``Im'' means the imaginary part. %The steady-state solution depends solely on the excitation frequency $\omega_s$. As to the transient part, it is given by the residues of the set of poles $\om_q$ of the transmission function $T(\om)$. 
The second contribution  $E_r(t)$ can be expressed as

\begin{equation} \label{eq:Transient}
E_{\textrm{r}}(t)= i \sum_{\{\omega_q\}} e^{-i \om_q t } \; 
\frac{\omega_s}{\om_q^2-\omega_s^2}\, E_0(\om_q) 
\left[ \frac{\partial T^{-1}} {\partial \om} \, (\om_q) \right]^{-1} \!\!\!.
\end{equation} 
The expression of the residues at $\om_q$ is given by
%---------------------- big equation
%\begin{equation} \label{eq:Qdiff}
%\begin{split}
%\frac{\partial T^{-1}} {\partial \om} \, (\om_q) = &
%-i \, \frac{1 + \ep_q}{2} \left[ 1 + \frac{\om_q}{2 \ep_q} 
%\frac{\partial \ep}{\partial \omega} (\om_q) \right] \cos[\om_q \sqrt{\ep_q}]
%- \\ 
%& \left[ \ep_q + \frac{2 \om_q \ep_q + i (\ep_q - 1)}{4 \ep_q}
%\frac{\partial \ep}{\partial \omega} (\om_q) \right] 
%\frac{\sin[\om_q \sqrt{\ep_q}]}{\sqrt{\ep_q}},
%\end{split}
%\end{equation}
\begin{widetext}
\begin{equation} \label{eq:Qdiff}
\frac{\partial T^{-1}} {\partial \om} \, (\om_q) =
-i \, \frac{1 + \ep_q}{2} \left[ 1 + \frac{\om_q}{2 \ep_q} 
\frac{\partial \ep}{\partial \omega} (\om_q) \right] \cos[\om_q \sqrt{\ep_q}]
- \left[ \ep_q + \frac{2 \om_q \ep_q + i (\ep_q - 1)}{4 \ep_q}
\frac{\partial \ep}{\partial \omega} (\om_q) \right] 
\frac{\sin[\om_q \sqrt{\ep_q}]}{\sqrt{\ep_q}} \,,
\end{equation}
\end{widetext}
%---------------------- big equation
where  $\ep_q \equiv \ep(\om_q)$. Each  term   in Eq. (\ref{eq:Transient}) represents 
the contribution of a pole $\omega_q$ to the transmitted field.
%\textcolor{\rev}{The exponential decay of each term depends on the imaginary part of the pole $\omega_q$}.

\section{\label{sec}field evaluation in time domain}

An elementary example is presented for a Drude-Lorentz medium with a dispersive permittivity  \cite{Ckittle}
\begin{equation}
\ep(\om) = 1 - \frac{\Om^2}{\om^2-\omr^2+i\om \gamma}  \, ,
\label{epDL}
\end{equation}
where $\Om$ is a coefficient of the medium related to the electron density, %\cite{Jackson}, 
$\omr$ is the resonance frequency of the dispersive medium, and $\gamma$ is the absorption coefficient.  All the aforementioned quantities are normalized, similar to (\ref{hom}). Numerical methods shall be used 
to compute the poles of $T(\omega)$ for complex systems \cite{Han03,Com03}, with the possibility to implement linearization techniques of the frequency dispersion \cite{Bru16,Fan10}.   In the present work,  Muller algorithm    is used for a numerical solution based on the secant method \cite{muller1956method}.{\revr{ In addition,  the simple  system considered here allows the derivation of asymptotic analytical  expressions for the poles,{\revrr{ as presented in the appendix.}} }}

Figure \ref{fig:LorentzPoles} shows the resonator poles $\{\omega_q\}$ of a given test case of $\omega_0=20$, $\Omega= 40$ for different values of the absorption coefficient: $\gamma= 0$, $0.1$, and  $0.5$.  For such a dispersive system, the imaginary part of the poles is frequency dependent: it  represents  the  losses due to absorption and the output coupling through the interfaces. % Therefore, each frequency component has a different decay rate at the output. {\revr{ This suggests that the high-frequency poles are predominant at the onset of the transmitted wave.}} 
{\revr{The poles in  Fig. \ref{fig:LorentzPoles} can be classified into two groups}}. The first group is  poles below resonance frequency $\om_0$, which are mainly responsible for the Brillouin precursor. The 
second group is poles above the plasma frequency $\omega_p = \sqrt{\om_0^2 + \Omega^2}$, 
i.e. $44.7$ in the given example. These high-frequency poles constitute  the wavefront of the signal with a velocity near $c$, i.e. the Sommerfeld precursor. No poles exist in between  since the permittivity is negative in this region. 
The absorption clearly affects the poles near both resonance and plasma frequencies.

%In the lossless case, where $\gamma \to 0$,  the imaginary part of the permittivity turns to be a Dirac function at $\pm \omr$, according to causality principle and the Kramers Kronig relations \cite{Jackson}. This Dirac contribution can be ignored because the transmission of the input interface of the slab vanishes as $\ep \to \infty$ at $\pm \omr$. %Taking the limit of the absorption term $\gamma \to 0$ gives the same results as putting $\gamma = 0$ in the Drude-Lorentz model. In this way, we can confidently use the Lorentz model for the lossless case. 

% Fig 2
%-------------------------------------------------------------------------------
\begin{figure}    
  \includegraphics[width=1\linewidth]{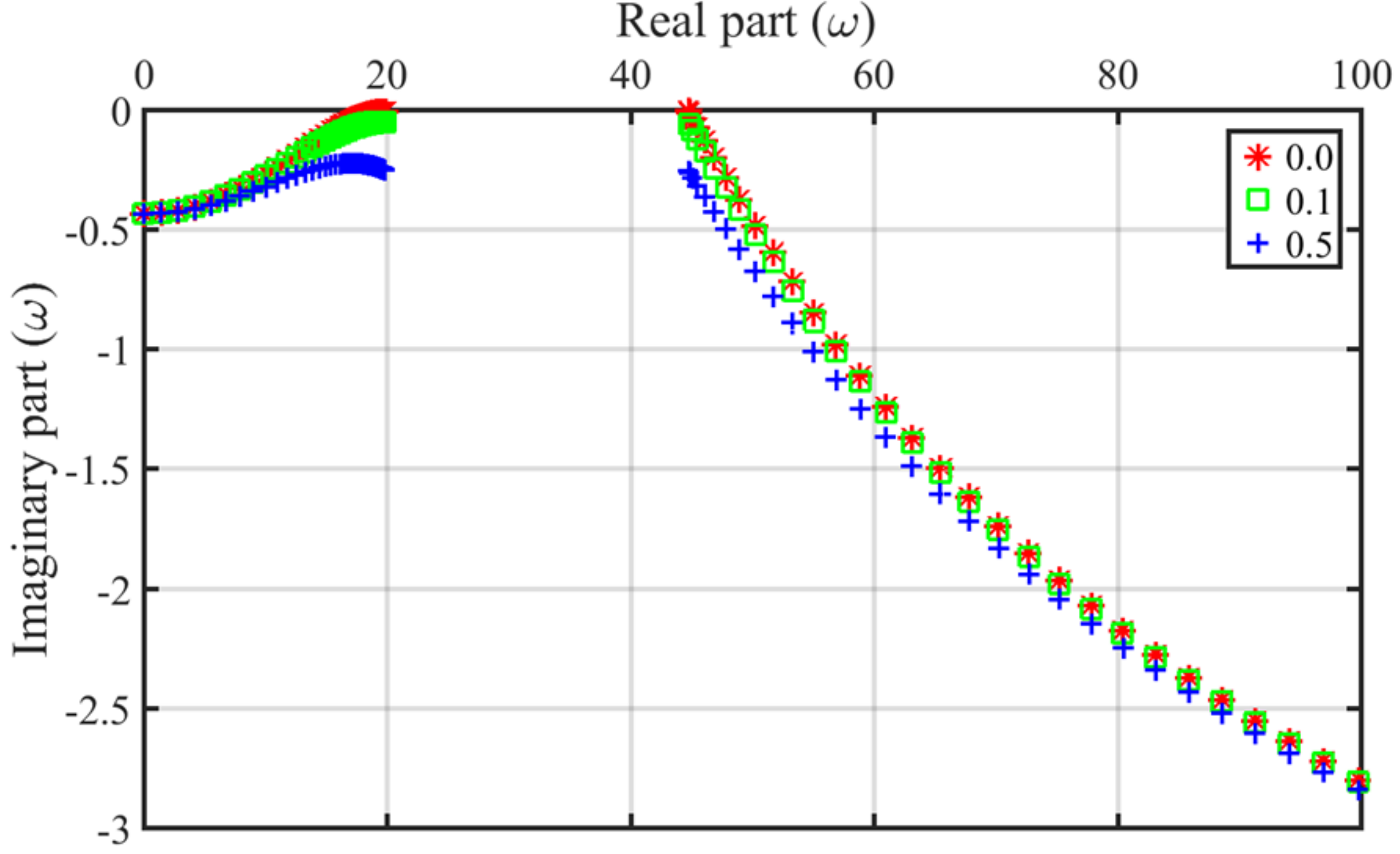}
 \caption[Optional caption]{The resonator poles \textcolor{\rev}{(modes)} of a Lorentz medium  of $\om_0 = 20$ and 
 $\Om = 40$ for the absorption coefficients $\gamma = 0$, $0.1$ and $0.5$.  The set of  poles is symmetric with respect to the imaginary axis.}
 \label{fig:LorentzPoles}
 \end{figure} 
%----------------------------------------------------------------------

\begin{figure} 
  \includegraphics[width=1\linewidth, keepaspectratio]{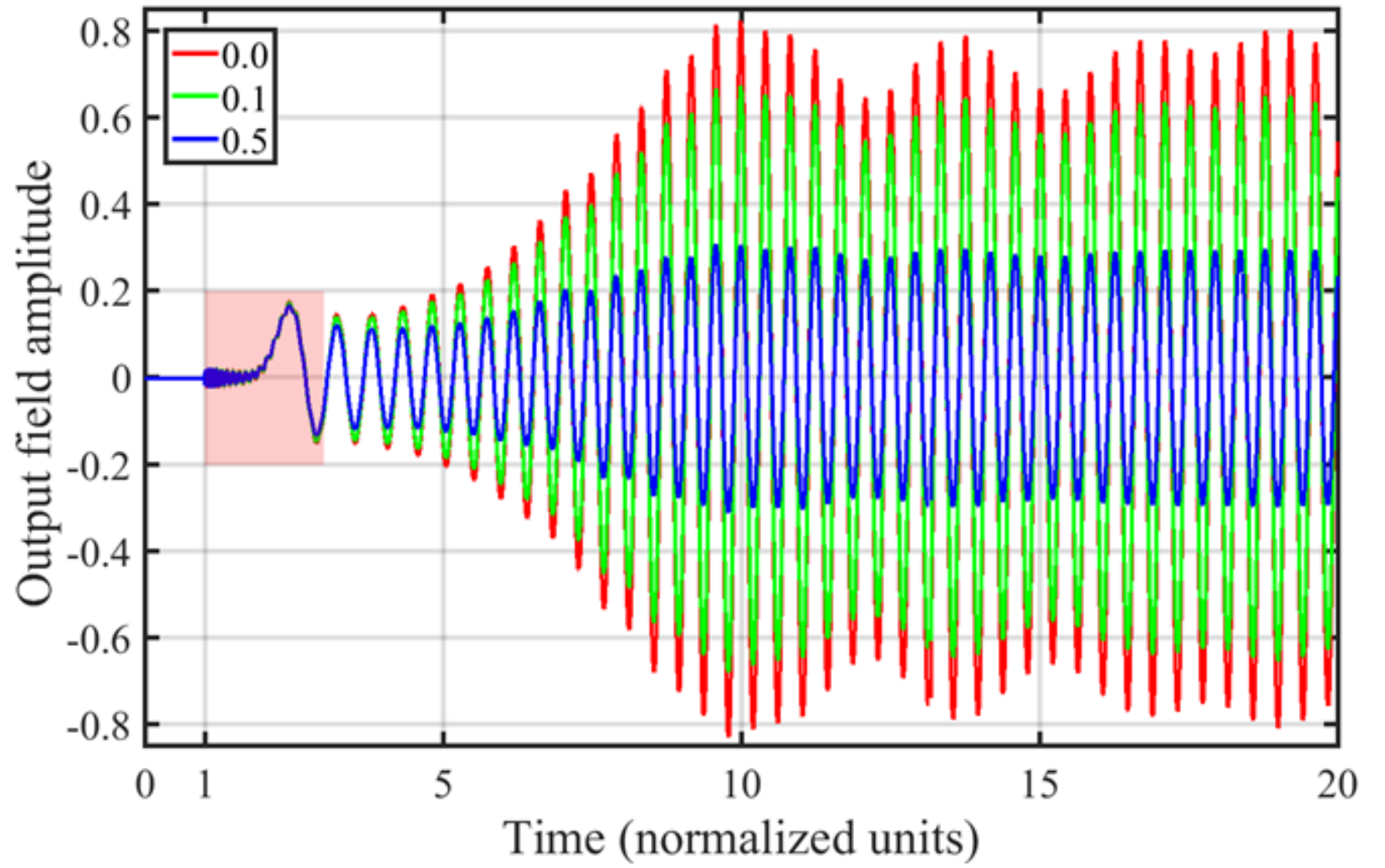}
 \caption[Optional caption]{The  temporal response of a Lorentz resonator of $\omega_0=20$,  $\Omega= 40$ and excited by a sinusoidal source at  $\omega_s=15$, switched on at $t=0$,  for the different absorption coefficients $\gamma=0$, $0.1$ and $0.5$.}
 \label{fig:2Main}
 \end{figure}

{The poles $\{\omega_q\}$ are then used to evaluate and interpret the transmitted field as given by Eqs. (\ref{twoparts}) to (\ref{eq:Qdiff})}. {\revrr{Figure \ref{fig:2Main} shows the  temporal evolution of the transmitted field $E_T(t)$  of the given test in   Fig. \ref{fig:LorentzPoles} for a near-resonance excitation of  $\omega_s=15$,  where the dispersion effect on the signal propagation is significant.}} The amplitude of the incident field is set to be unity at the input of the slab. {\revrr{As depicted in Fig. \ref{fig:2Main}, the  signal}} shows successive various temporal regimes since each frequency component of the signal can be considered to have its own velocity. At the onset of the signal {\revrr{(highlighted region)}}, the high frequency but low amplitude oscillations correspond to the Sommerfeld precursor, followed by the low-frequency oscillations constituting the Brillouin precursor. {\revrr{Afterward, the main signal starts to build up at the output, until it reaches the steady level $|T(\omega_s)|$ after few round trips inside the resonator. The solution accuracy  depends on the precision of the poles. The effect of absorption on the precursors  is negligible since both Sommerfeld and Brillouin poles are far away from the resonance.}}

\section{Precursors}

\begin{figure}[t!]
\includegraphics[width=1\linewidth, keepaspectratio]{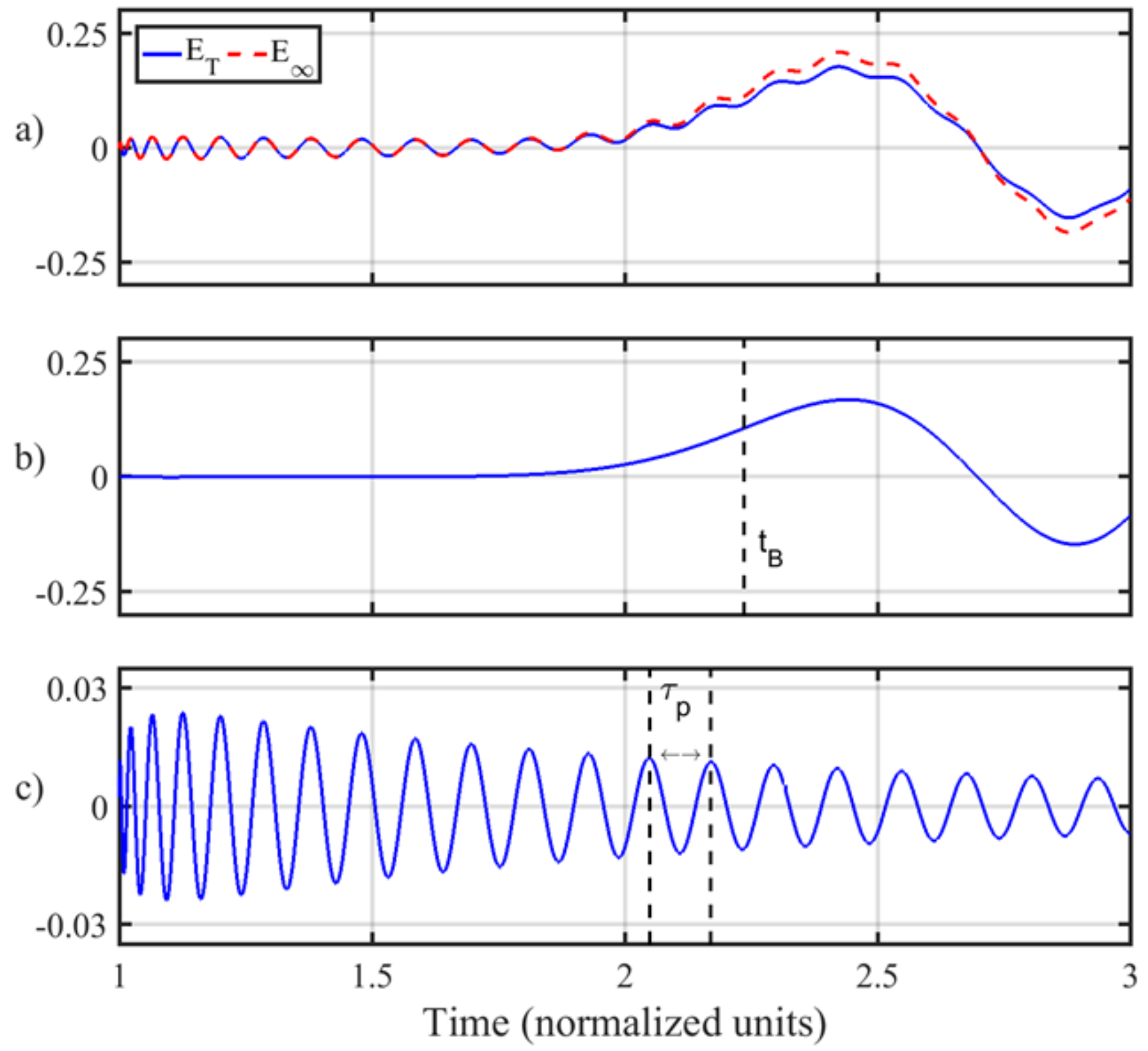}
\caption{ a) The total transmitted signal $E_T$ for the case of a slab [blue curve, Eq. (\ref{intslab})]{\revrr{ at the highlighted region in Fig.~\ref{fig:2Main}}} showing the Sommerfeld precursor starting at $t=1$  followed by the Brillouin precursor,{\revrr{ compared to the infinite medium solution $E_\infty$}}
[red dashed, Eq. (1)]. b) and c) The  contributions of the 
low and high-frequency groups of poles.}
\label{Fig3}
\end{figure}

The main advantage of the  modal approach is the direct link between the poles shown in  Fig.  \ref{fig:LorentzPoles}, and the different components of the precursors shown in {\revrr{Fig.  \ref{Fig3}.a): the highlighted region at Fig. \ref{fig:2Main}. }} {\revr{For instance,  the curve in \ref{Fig3}.b)  illustrates the sole contribution of  low-frequency poles that is shown to be responsible for the Brillouin precursor. }} The arrival time of
this Brillouin precursor peak $t_B$ can be directly evaluated using the group velocity approximation since, in the low-frequency regime, the dispersion is weak. 
As confirmed by the curve \ref{Fig3}.b), this time $t_B$ is approximately 
given by$\sqrt{\ep(\omega=0)}\approx2.24$.  The contribution 
of the poles near resonance is negligible since the associated reflectivity at the interfaces is close to unity{\revrr{, refer to the appendix for a detailed discussion.}}{\revb{
 Figure \ref{Fig3}.c) shows the contribution of the {\revb{high-frequency}} 
poles above the plasma frequency $\omega_p$: it starts with the Sommerfeld 
precursor,  corresponding to the highest frequencies, 
then followed by the frequencies {\revr{gradually decreasing down to} $\omega_p$, as confirmed by the 
temporal period $\tau_p = 2 \pi / \om_p \approx 0.14$ for $t \ge 2$. }}}

{\revb{{\revr{It is important to notice that this}} method is applicable to accurately describe the 
transient waves propagating in an infinite dispersive medium. Indeed, the curves derived from 
Eqs. (\ref{eq:1maineq}) and (\ref{intslab}) appear to be indiscernible at the onset of the signal, i.e. 
for the wave components with group velocity close to $c$, as clearly shown in Fig. \ref{Fig3}.a). 
That motivates the derivation of the Sommerfeld precursor using the proposed method 
given that only the high-frequency poles above $\omega_p$ contributes, 
as {\revr{confirmed}} by Figs. \ref{Fig3}.b) and \ref{Fig3}.c).}}
%-------------------------------------------------------------------------------
% Fig 4
%-------------------------------------------------------------------------------
%-------------------------------------------------------------------------------
An estimate of the corresponding poles above $\omega_p$ is provided in the {\revrr{appendix, Eq. 
(\ref{eq:zone4})}}: for integers $q \in \mathbb{Z}$,  
\begin{equation}\label{eq:omq}
\om_q \approx q \pi {\revb{ \left[ 1+\frac{\Omega^2 / 2}{q^2 \pi^2} \right] }} 
- i \ln \dfrac{4 q^2 \pi^2 }{\Omega^2} 
%\, , \qquad |q| \pi \ge  \om_p 
\, .
\end{equation} 
The corresponding residues are determined using  Eq. (\ref{eq:Qdiff}) which, in this 
situation of high frequencies, gives $i \exp[i \om_q \sqrt{\ep_q}] /(2 \sqrt{\ep_q})$ 
at the same order of approximation.
Assuming $\om_q \gg \om_s$ and $\om_q \gg \om_0$,  Eq. (\ref{eq:Transient}) 
can be used to obtain an estimate of $E_p(t)$, the contribution of the high-frequency poles:
\begin{equation} \label{eq:ES}
E_{p}(t) \, {\revb{\approx}} \, {\revb{- \omega_s}} \sum_{{\revb{q}}} 
{\revb{ \frac{\exp[-i \om_q t]}{\om_q^2} }} \;{\revb{\frac{\exp[i \omega_q \sqrt{\ep_q}\,] }{2 \sqrt{\ep_q}}}}. 
\end{equation} 
Using the asymptotic expression of the poles (\ref{eq:omq}) and the  first order estimate
$\sqrt{\ep_q} \approx 1 - \Omega^2 / (2\om_q^2)$, the expression 
above becomes 
\begin{equation} \label{eq:ESpoles}
E_{p}(\tau) \, \approx \, - {\revb{\frac{\omega_s}{2} }} \, 
 \sum_{{\revb{q}}}  
%\dfrac{1}{(q \pi)^{2+2\tau}} 
%\exp\left[ -i q \pi \tau - i \dfrac{\Omega^2}{2 q \pi} \right] , 
{\revb{\frac{e^{-i [ q \pi \tau + \Omega^2 (\tau+1)/(2 q \pi )] } }{(q \pi)^{2} \: (2 q \pi / \Omega)^{2\tau}} \, 
\left[ 1+\frac{\Omega^2 / 2}{q^2 \pi^2} \right]}},
\end{equation} 
where $\tau = t - t_0$ ($t - 1$ in normalized units) 
is the relative time. {\revb{All the poles corresponding to $q \in \mathbb{Z}$ are considered, and then{\revrr{ the discrete sum can be directly translated to the integral,
\begin{equation}
E_{p}(\tau) \approx - \dfrac{\omega_s}{2 \pi}  \, 
\mathcal{P} \displaystyle\int d\om \, 
\frac{e^{ -i [ \om \tau + \Omega^2 (1 + \tau)/(2 \om)] } }{\om^{2} \: (2 \om / \Omega)^{2 \tau}} 
\left[ 1+\frac{\Omega^2 / 2}{\om^2} \right] , 
\label{supp:prec}
\end{equation}
where the symbol $\mathcal{P}$ means that the Cauchy principal value of the integral
is considered. It can be checked that, for $\tau>0$, the integral over the semi-circle 
$C_R^-$ with radius $R \rightarrow \infty$ in the lower part of the complex plane of frequencies (with negative imaginary part) vanishes. Also, it can be shown that the same 
integral over the semi-circle $C_\rho^+$ with radius $\rho$ tending to zero in the upper part of the complex plane of frequencies (with positive imaginary part) tends to zero.
Indeed, let $\om=\rho \exp[i\phi]$ and consider the integral in Eq. (\ref{supp:prec}) on the path with $\phi$ varying from $\sqrt{\rho}$ to $\pi - \sqrt{\rho}$. Then, for $\rho \ll 1$ the integral is bounded by 
\begin{equation}
\displaystyle\int_0^\pi d\phi\, 
\frac{e^{ [ \rho \tau - \Omega^2 (1 + \tau)/(2 \sqrt{\rho})] } }{\rho \: (2 \rho / \Omega)^{2 \tau}} 
\left[ 1+\frac{\Omega^2 / 2}{\rho^2} \right] \: \underset{\rho\rightarrow 0}{\longrightarrow} \: 0 \, .
\label{supp:prec2}
\end{equation}
Hence, adding the vanishing integrals along $C_\infty^-$ and $C_0^+$, the integral expression can be written as
\begin{equation}
E_{p}(\tau) \approx - \dfrac{\omega_s}{2 \pi}  \, 
\displaystyle\int_{C} d\om \, 
\frac{e^{ -i [ \om \tau + \Omega^2 (1 + \tau)/(2 \om)] } }{\om^{2} \: (2 \om / \Omega)^{2 \tau}} 
\left[ 1+\frac{\Omega^2 / 2}{\om^2} \right] , 
\label{supp:prec3}
\end{equation}
where $C$ is a closed loop around the origin $\om = 0$ and negatively oriented. This loop can be deformed, as soon as the origin remains inside, since the function under the integral in Eq. (\ref{supp:prec}) is analytic in $\mathbb{C} \setminus \{ 0 \}$. Thus, taking $C = U$, the circle centered at the origin and positively oriented, the following expression  of the precursor is obtained,}}
\begin{equation} \label{eq:ES2}
E_{p}(\tau) \approx {\revb{\dfrac{\omega_s}{2 \pi} }} \, 
\displaystyle\int_{U} d\om \, 
{\revb{ \frac{e^{ -i [ \om \tau + \Omega^2 (1 + \tau)/(2 \om)] } }{\om^{2} \: (2 \om / \Omega)^{2 \tau}} 
\left[ 1+\frac{\Omega^2 / 2}{\om^2} \right]. }}  
\end{equation} 

{\revb{This integral expression is similar to the one proposed by 
Sommerfeld in Ref. \cite{brillouin1960wave}, but with three differences. The 
first difference is the correction $\tau$ in the term $\Omega^2 (1 + \tau)/(2\om)$ in the 
exponential function, which induces a correction in the oscillation period of the 
Sommerfeld precursor. The second difference is the time-dependent correction factor 
$(2 \om / \Omega)^{2\tau}$ decreasing the amplitude of the 
Sommerfeld precursor. Finally, the third difference is the additional 
term on the right with the factor $\Omega^2/(2 \om^2)$ which allows the definition of 
the precursor after the one of Sommerfeld and before the one of Brillouin.
These three corrections lead to significant improvements in the description 
of the precursor amplitude and periodicity, as shown hereafter.  

%%%%%%%%%%%%%%%%%%% for supplemental
%The expression (\ref{eq:ES2}) can be evaluated remarking that the same integral on the path 
%$\overline{\Gamma}$, the line parallel to the real axis in the lower half plane 
%of complex frequencies with negative imaginary part, vanishes for $\tau>0$ since the path 
%$\overline{\Gamma}$ can be deformed into a semi-circle with infinite radius in the lower 
%half plane. Substracting this integral over $\overline{\Gamma}$ to the one in (\ref{eq:ES2}), 
%it is obtained that  the integrals 
%of the terms with $\exp[\pm i \om \tau]$ over  a semi-circle with infinite radius in the 
%half planes $\pm\text{Im} \om >0$ vanish. 
%%%%%%%%%%%%%%%%%%%%%%%%%%
The expression (\ref{eq:ES2}) can be evaluated by choosing the {\revr{frequencies $\omega = \nu_\tau \,e^{i \phi}$ }} on the circle $U$ with radius 
$\nu_\tau$ defined by}}
\begin{equation}
{\revb{\nu_\tau = \Omega \: \sqrt{ \dfrac{1 + \tau}{2 \tau}} \, .}}
\end{equation} 
{\revb{ The following expression is then obtained}}
{\revb{
\begin{equation}
\label{eq:ESint2}
\begin{array}{lr}
E_{p}(\tau) \hspace*{-1mm} & \approx  \dfrac{2 \omega_s}{\Omega} \,
\left[ \dfrac{\Omega}{2 \nu_\tau} \right]^{1+2\tau} \!  \dfrac{i}{2\pi} 
\displaystyle\int_{0}^{2\pi} \hspace*{-3mm} d\phi \,
e^{i [ -(1+2\tau) \phi  - 2 \nu_\tau \tau \cos\phi]} \hspace*{1.7mm} \\[4mm]
& + \, \dfrac{4\omega_s}{\Omega} \,
\left[ \dfrac{\Omega}{2 \nu_\tau} \right]^{3+2\tau} \! \dfrac{i}{2\pi} 
\displaystyle\int_{0}^{2\pi} \hspace*{-3mm} d\phi \,
e^{i [ -(3+2\tau) \phi  - 2 \nu_\tau \tau \cos\phi]} \, .
\end{array}
\end{equation} 
The integrals above are close to the Bessel functions of irrational order 
$1+2\tau$ and $3+2\tau$.}} 
Hence, the final obtained expression for the precursor is
{\revb{
\begin{equation}
\label{eq:precursor}
\begin{array}{lr}
E_{p}(\tau) \hspace*{-1mm} & \approx  \dfrac{2 \omega_s}{\Omega} \,
\left[ \dfrac{\Omega}{2 \nu_\tau} \right]^{1+2\tau} J_{1+2\tau}\big[ 2 \nu_\tau \, \tau \big] \hspace*{1.7mm} \\[4mm]
& - \, \dfrac{4\omega_s}{\Omega} \,
\left[ \dfrac{\Omega}{2 \nu_\tau} \right]^{3+2\tau}  J_{3+2\tau}\big[ 2 \nu_\tau \, \tau\big]  \, .
\end{array}
%E_{p}(\tau) \approx  \dfrac{2 \omega_s}{\Omega} \,
%\left[ \dfrac{\Omega}{\nu_\tau} \right]^{1+2\tau} \hspace*{-5mm} 
%J_{1+2\tau}\big[ \nu_\tau \, \tau \big] - \, \dfrac{\omega_s}{\Omega} \,
%\left[ \dfrac{\Omega}{\nu_\tau} \right]^{3+2\tau}\hspace*{-5mm} 
%J_{3+2\tau} \big[ \nu_\tau \, \tau\big]  \, .
\end{equation} 
This expression converges to the one obtained by Sommerfeld \cite{brillouin1960wave} 
when $\tau \to 0$.}}  Figure \ref{fig:SommerComp} shows the comparison between 
the different analytical estimates of the precursors and the numerical 
temporal response for the given test case in Fig. \ref{Fig3}.a). It turns out that the new expression (\ref{eq:precursor}) 
obtained by the present method is significantly more accurate than the original one 
proposed by Sommerfeld. {\revb{In particular, in Fig. \ref{fig:SommerComp},
both the amplitude and oscillating frequency of the signal remain well described 
by the analytical estimate (\ref{eq:precursor}) until the arrival of the Brillouin precursor at 
$1 + \tau \approx 2$. The oscillating frequency, defined by 
\begin{equation}
\dfrac{\partial \, 2 \nu_\tau \tau}{\partial \tau} = \Omega \, \dfrac{1+2\tau}{\sqrt{2\tau(1+\tau)}} \, ,
\label{freq}
\end{equation}
tends to infinity when $\tau \rightarrow 0$ and, for large values of $\tau$, 
decreases to $\Omega \sqrt{2}$. This asymptotic value $\Omega \sqrt{2}$ is 
different from the expected frequency $\om_p$ [see Fig. \ref{Fig3}.c)], which can 
be explained by the absence of the poles in the vicinity of $\om_p$ in the 
starting expression 
(\ref{eq:ES}).}}

{\revb{Finally, the relevance of the group velocity for the precursor is discussed. Starting 
from the {\revr{normalized}} dispersion law $ k^2 \approx \om^2 [1 - \Omega^2/(\om^2-\om_0^2)]$, with {\revr{$k d \rightarrow k $ 
the normalized }} wavenumber, the group velocity can be estimated for high frequencies: $\partial \om/\partial k 
\approx  [ 1 - \Omega^2/(2 \om^2)]$. Assuming that the waves travel 
at this group velocity, the time needed to propagate along the distance $d$ is given by 
\begin{equation}
(1 + \tau) \dfrac{\partial \omega}{\partial k} = 1 \quad \Longleftrightarrow \quad 
\om_g(\tau) = \Omega \sqrt{\dfrac{1+\tau}{2 \tau}} \, .
\label{gv}
\end{equation}
Hence, it is found that the frequency $\om_g(\tau)$ is presicely the radius $\nu_\tau$ of 
the circle $U$ used in the derivation of the precursor, showing the critical role of this 
frequency {\revr{corresponding to}} the group velocity. Moreover, when 
$\tau \rightarrow 0$, the frequency $\om_g(\tau) \sim \Omega / \sqrt{2\tau}$ tends to 
the oscillation frequency (\ref{freq}). However, for values of time $\tau > 0$, the ocillating 
frequency (\ref{freq}) is different from $\om_g(\tau)$, showing that the velocity of {\revr{each frequency component }} appears to be more complex than the group velocity, even in this situation 
of low dispersion occuring at high frequencies. 
}}
%-------------------------------------------------------------------------------
% Fig 5
%-------------------------------------------------------------------------------

%-------------------------------------------------------------------------------

\begin{figure} [t!]
\includegraphics[width=1\linewidth]{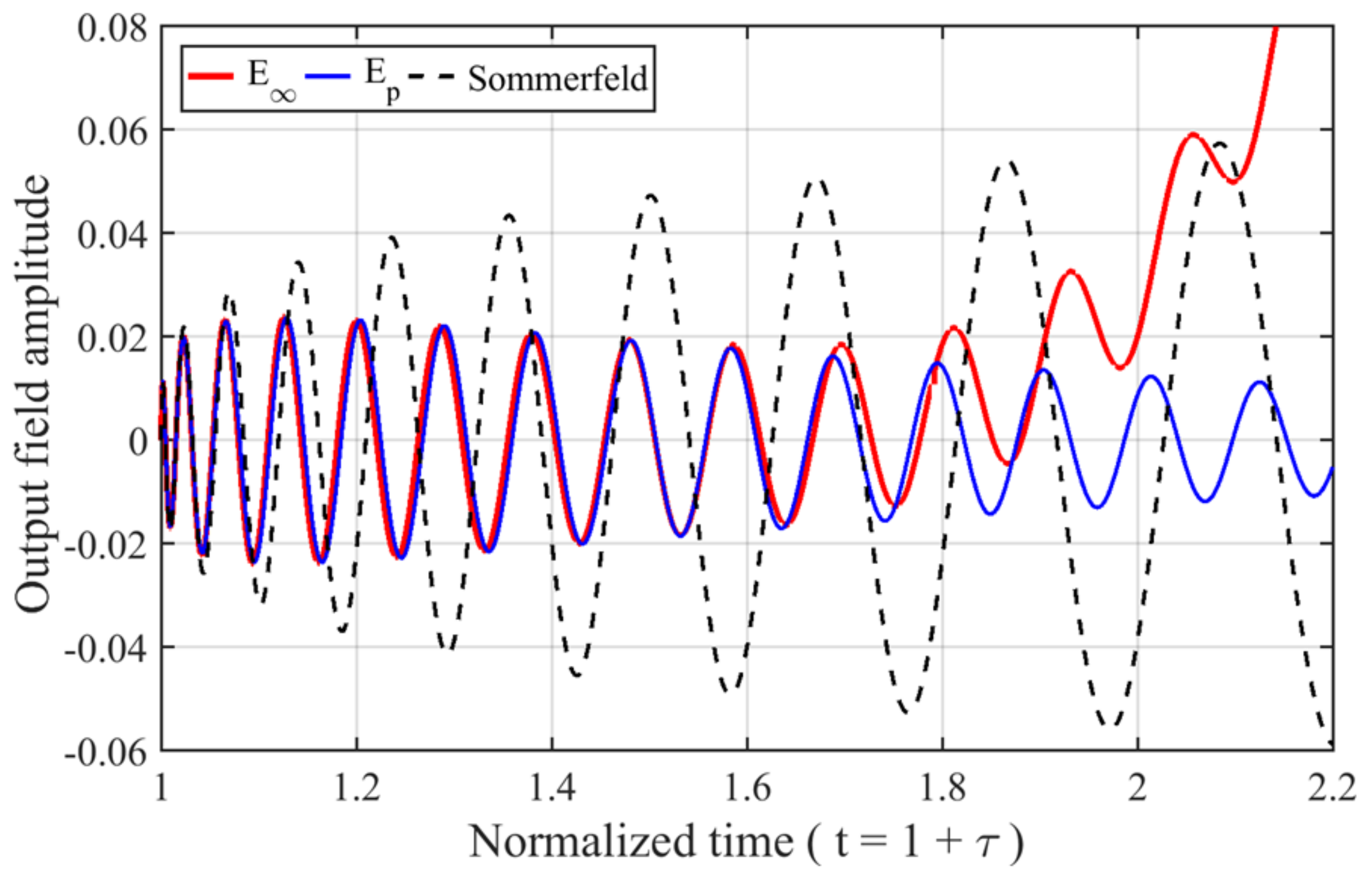}
\caption[Optional caption]{ {\revb{A comparison between the  precursors of a wave propagating in an infinite  medium $E_\infty(t)$ given by (1) (in red), the  Sommerfeld precursor original expression 
\cite{brillouin1960wave} (in dashed black), and the precursor 
$E_p(\tau)$ given by the novel expression (\ref{eq:precursor}) (in blue).}} }
\label{fig:SommerComp}
\end{figure}

\section{conclusion}

{\revb{In conclusion, a novel method has been introduced to analyze the wave 
propagation in dispersive media. The dispersive medium is considered to be {\revr{bound}}, which 
generates a discrete set of complex resonances and modes. {\revr{In this case}}, it is shown that 
no branch-cut exits in the complex frequency domain. Consequently, a closed-form expression 
of the transmitted field is derived using the modal expansion. That allows a direct physical  interpretation of different components of the transient field as different modes of the system, while these transients  are characterized in terms of the oscillation frequency and arrival time. 
 In addition, this method can effectively describe the precursors 
in infinite dispersive media. A modified analytical expression of the Sommerfeld 
precursor is derived, which significantly improves the description of the precursor 
amplitude and oscillation period. In particular, the obtained accuracy for the first 
oscillations  {\revr{can be exploited}} to design early detection systems for waves.}}

The present work shows the relevance of the method using an expansion of the modes of 
the system and, more generally, the potential of the quasinormal modes expansion 
\cite{Sau13, Bai13, Vial14}. The proposed method and results are 
valid for all waves governed by  Helmholtz equations with dispersion subject to 
causality requirement, e.g. elastic, hydrodynamic, and gravity waves.

\begin{acknowledgments}
{\revb{This work was supported by the French National Agency for Research (ANR) 
under the project ``Resonance'' (ANR–-16–-CE24–-0013).}} We would like to express our gratitude to Prof. Aladin Hassan Kamel (Ain-Shams University, Egypt) for the valuable discussions. {\revb{We acknowledge the anonymous reviewers for their  valuable comments.}}

%{\revb{This work was supported by the French National Agency for Research (ANR) under the project ``Resonance'' (ANR–-16–-CE24–-0013).}} We acknowledge Prof. Aladin Hassan Kamel (Ain-Shams University, Egypt) and the anonymous reviewers for their  valuable comments.
 
 \end{acknowledgments}

\renewcommand{\theequation}{A.\arabic{equation}}
\setcounter{equation}{0}

\renewcommand\thefigure{A.\arabic{figure}}    
\setcounter{figure}{0} 
 \appendix
\section{Analytic treatment of the poles of the Lorentz medium}

In order to evaluate the temporal response of dispersive media, it is required to obtain the set of poles 
$\{\om_q, q \text{ integer}\}$ of the resonator transmission function $T(\omega$) in Eq. (\ref{eqSlabTF}), or in other words, 
to evaluate all the solutions of $\om_q$ for
\begin{equation}
1-\rho^2(\om_q) \, e^{2 i \om_q \sqrt{\ep(\om_q)}} = 0 \, .
\end{equation}
In this appendix, an asymptotic approach is proposed 
in order to estimate the poles for the specific case of Lorentz medium 
with the frequency-dependent permittivity expression:
\begin{equation}
\ep(\om) = 1 - \frac{\Om^2}{\om^2-\omr^2+i\om \gamma} \, . 
\label{epDL}
\end{equation}
This could provide more understanding to the contribution to the temporal response of 
 different types of poles. 
Notice that the complex frequency domain is symmetric around the 
imaginary axis,  i.e. $\text{Re}(\om_{-q})=-\text{Re}(\om_{q})$, and 
$\text{Im}(\om_{-q})=\text{Im}(\om_{q})$. 

The solution is determined in the form $\om_q= a(q)-ib(q)$, 
where a and b are positive real parameters that represent the 
Free Spectral Range (FSR) and the losses of the resonator at a given frequency, respectively.  
With these notations, the equation above becomes
\begin{equation} \label{eq:MainApp}
   \rho(\om_q) = \pm \; e^{-i \omega_q \sqrt{\ep(\om_q)}}.
\end{equation}
Assuming lossless media $ \gamma=0$, equating the amplitude and the phase parts leads to
\begin{equation}
a(q)= \frac{q \pi}{\sqrt{\ep(\om_q)}},
\label{A:a}
\end{equation}
\begin{equation}
b(q)= - \frac{\ln \big| \rho(\om_q) \big|}{\sqrt{\ep(\om_q)}}.
\label{A:b}
\end{equation}
Figure \ref{fig:epsallzones} shows four different zones of the permittivity of a Lorentz medium,  
where the poles at each zone can be characterized by a distinct feature. Two zones are below resonance $\om_0$ and two are above the plasma frequency 
$\omega_p = \sqrt{\omr^2 + \Omega^2}$. No poles exist  between $\omr$ and $\om_p$ where the permittivity  is negative and the index is purely imaginary. This leads to a decay of the signal inside the medium, so there is no propagation solution.

The first zone is at low frequencies, where the permittivity is almost constant. The poles in this zone constitute the Brillouin precursor. The second zone is near resonance frequency, where the permittivity  is highly dispersive.  The third zone is near the plasma frequency, $\omega_p = 44.7$ in the given example,  where $ \ep \approx 0$. The fourth zone is at high frequencies. \textcolor{\rev}{ The high-frequency poles in both zones 3 and 4 are is responsible for the Sommerfeld precursor at the onset of the propagating wave.}

At each frequency zone in Fig. \ref{fig:epsallzones}, a reasonable approximation 
can be used to formulate a closed-form expression of the poles. The poles between these zones can be determined using an iterative method. 
Since the amplitude term (reflectivity) is bounded to unity, it is generally correct to assume 
that $b(q) \ll a(q)$ and $|\om_q| \simeq a$.%, except at the static value $q=0$. 

\begin{figure}    
  \includegraphics[width=1\linewidth, keepaspectratio]{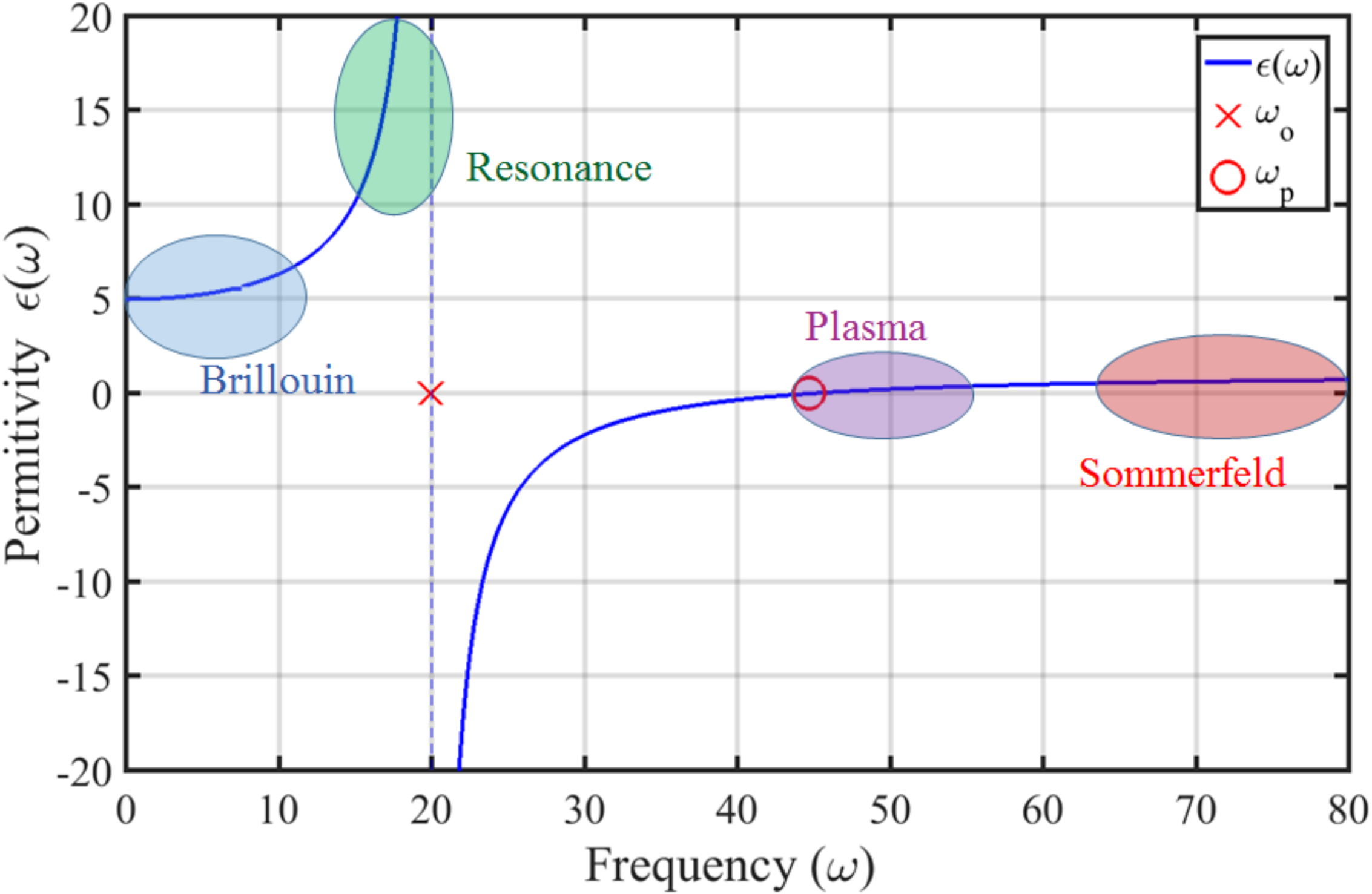}
 \caption[Optional caption]{The permittivity of a Lorentz medium of $\om_0 = 20 $, $\Om=40$, \textcolor{\rev}{and $\gamma=0$}. Four distinct zones of poles are indicated.}
 \label{fig:epsallzones}
 \end{figure}
%------------------------------------------------------------------------------------------------
\subsubsection{[Zone-1]: Low-frequency (Brillouin precursor) } \label{sec:DCzone}
%------------------------------------------------------------------------------------------------
Starting from $\om \ll \omr$, Taylor series can be used at this zone to approximate the permittivity, and the reflectivity in Eq. (\ref{eq:r})  as
\begin{equation} \label{eq:zone1eps}
\ep(\om)  \underset{\om \ll \omr}{\approx}  \eps + (\eps - 1 ) \, \dfrac{\om^2}{\omr^2},
\end{equation} 
\begin{equation}
\begin{array}{ll}
\rho(\om) & \approx \rho_s \left[1 + \dfrac{1}{\sqrt{\eps}} \, \dfrac{\om^2}{\omr^2} \right],
\end{array}
\end{equation} 
where $\eps = 1 + \Om^2 / \omr^2$ is the static permittivity at $\om = 0 $, and $ \rho_s = \rho(\om=0)$.
Since the permittivity is nearly constant in this zone, the real part of the poles can be assumed to be
\begin{equation}
a(q) = \frac{q\pi }{\sqrt{\ep_s}} \, .
\end{equation}
Using $|\om_q| \simeq a$, % for $ q > 0$,
it is obtained that 
\begin{equation}
b(q) = \dfrac{\ln \big| 1/\rho_{\om_s} \big|}{\sqrt{\eps}} - \dfrac{q^2 \pi^2}{ \omr^2 \eps} \, .
\end{equation}
Consequently, the poles expression for zone 1 is
\begin{equation} \label{eq:zone1}
\omega_q = \frac{q\pi }{\sqrt{\ep_s}} - i \dfrac{\ln \big| 1/\rho_{\om_s} \big|}{\sqrt{\eps}}
+ i \dfrac{q^2 \pi^2}{\omr^2 \eps} \, .
\end{equation}
This expression is valid as long as $q  \ll  \sqrt{\ep_s} \omr$.

Since the permittivity is weakly dispersive in this zone, it is possible to use the group 
velocity approximation in order to estimate the arrival time of the Brillouin precursor 
(in normalized units) as,
\begin{equation}
t_{B} = \sqrt{\epsilon_s}.
\end{equation}
%------------------------------------------------------------------------------------------------
\subsubsection{ [Zone-2]: Near-Resonance  ($\ep \to \infty$)}
%------------------------------------------------------------------------------------------------
In the near resonance zone $\omega \simeq \omr$, the permittivity  can be approached by
\begin{equation} \label{epr}
\ep(\om)  \underset{\om \to \omr}{\approx}  \dfrac{\Om^2}{2 \omr} \, \dfrac{1}{\omr - \om} \, . 
\end{equation} 
The poles are found using an iterative method. As a first step, let $\omega_q=\omr - \zeta^2$, which implies 
\begin{equation} \label{eq:Ressqrteps} 
\sqrt{\ep(\omega_q)} \simeq   \frac{ \Omega}{\sqrt{2\omr} \sqrt{(\omr-\omega_q)}}  \simeq   
\frac{ \Omega}{\sqrt{2\omr}} \frac{1}{\zeta} \, ,
\end{equation}
and, assuming $\zeta^2 \ll \frac{\Omega^2}{2\omr}$, it is obtained that
\begin{equation}
\rho(\omega_q) \simeq -1 + \frac{2}{\sqrt{\epsilon(\om_q)}} = - 1 + 2 \frac{\sqrt{2\omr}}{\Omega} \, \zeta \, .
\end{equation}
%The positive root is chosen to satisfy the causality of the fourier transform, 
%(convergance of for $t>0$). 
Taking the logarithm of the square of Eq. (\ref{eq:MainApp}) 
and using that $\ln(1+x) \approx x$ for $x \ll 1$ yield
\begin{equation}
- 4 \frac{\sqrt{2\omr}}{\Omega} \zeta = i q 2 \pi - i 2 \dfrac{ \Omega}{\sqrt{2\omr}} \, \dfrac{1}{\zeta}\, 
(\omr-\zeta^2) \, .
\end{equation}
Since $\zeta^2  \ll \frac{\Omega^2}{2\omr}$, the left side can be omitted. 
In order to reach an expression for $\zeta$ for the first iteration, it is assumed 
in addition that $ \zeta^2 \ll \omr$, which leads to
\begin{equation}\
\zeta=   \frac{ \Omega \sqrt{\omr / 2}   }  {  q \pi }\, .
\end{equation}
Consequently, the poles expression is, after a first iteration, 
\begin{equation}\
\omega_{q} = \omr -  \frac{ \omr  \Omega^2 }  { 2 \pi^2 q^2  }, 
\end{equation}
As expected, no pole exists for frequencies higher than the resonance 
where the refractive index is complex (metallic region).

To obtain the imaginary part of the poles expression, a second iteration is needed: 
let now $\om_q$ be
\begin{equation} \label{eq:respoleassm}
\omega_{q} = \omr -  \frac{ \omr  \Omega^2 }  { 2 \pi^2 q^2  } (1+\zeta^2) \, .
\end{equation}
Here, it is assumed that $\zeta^2 \ll 1$. Similarly to the first 
iteration, we have
\begin{equation}\
\ep(\om_q)\simeq   \frac{\pi^2 q^2}{ \omr^2 (1+\zeta^2)} \, ,
\end{equation}
and
\begin{equation}\  \label{eq:rhoResN}
\rho(\om_q)  = - 1 +  \frac{2\omr}{q \pi} \left[ 1+\frac{\zeta^2}{2} \right] \, .
\end{equation}
Taking again the logarithm of the square of Eq. (\ref{eq:MainApp}), 
it is obtained that
\begin{equation}
\begin{array}{ll}
- \dfrac{4\omr}{q \pi} \left[ 1+\dfrac{\zeta^2}{2} \right] = \\[4mm]
\qquad \qquad i q 2\pi - i \dfrac{ 2\pi q}{\omr} 
 \left[ 1-\dfrac{\zeta^2}{2} \right]  \left[ \omr -  
 \dfrac{ \omr  \Omega^2 }  { 2 \pi^2 q^2  } (1+\zeta^2) \right] \, .
 \end{array}
\end{equation}
Since $q \gg  \Omega$ and $\zeta^2 \ll 1$, this implies
\begin{equation}\label{eq:zetares}
- \dfrac{4\omr}{q \pi} \approx i q \pi \zeta^2
\quad \Longrightarrow \quad \zeta^2 = i \frac{4 \omr}{q^2\pi^2}.
\end{equation}
The expression of poles obtained after the two iterations is then

\begin{equation}\ \label{nearres}
\omega_{q} = \omr -  \frac{\omr  \Omega^2 }  {2  q^2 \pi^2 } \left[ 1 +i \frac{4\omr}{q^2\pi^2} \right] \, .
\end{equation}

This expression is limited to $q \gg  (\sqrt{\omega_0},\Omega)$.

The poles in this zone show an accumulation behavior, i.e. an  infinite number of poles 
in the zone near the resonance frequency. However, one can show that the summation of the overall 
contribution of these infinite poles, in Eq. (\ref{eq:Transient}), converges. That's due to the fact that by 
approaching the resonance frequency, the reflectivity of the interfaces goes to unity owing 
to the very large permittivity mismatch. Therefore, the transmission function of the 
resonator vanishes near resonance $T(\omega \to \omr)\to 0$.  
This can by shown by analyzing the term corresponding to the transmission function of the 
resonator in  Eq. (\ref{eq:Transient}),
\begin{equation} \label{eq:AQdiff}
\begin{split}
\frac{\partial T^{-1}} {\partial \om} \, (\om_q)   \underset{\om_q \approx \om_0}{\approx}  &
-i \, \left[  \frac{\om_0}{4} 
\frac{\partial \ep}{\partial \omega} (\om_q) \right] \cos[\om_0 \sqrt{\epsilon_q}]
- \\ 
& \left[ \epsilon_q + \frac{2 \om_0 + i }{4 }
\frac{\partial \ep}{\partial \omega} (\om_q) \right] 
\frac{\sin[\om_0 \sqrt{\epsilon_q}]}{\sqrt{\epsilon_q}},
\end{split}
\end{equation}
The term $ [\partial \ep/\partial \omega] (\om_q)$ depends on  $1/(\omr-\om_q)^2$,
which in turn, depends on $q^4$. The limit of Eq. (\ref{eq:Qdiff}), for
$ q \to \infty$,  tends to $\infty$. 
Hence, $ [(\partial T^{-1}) / (\partial \om) ] ^{-1} \to 0 $. That means the higher 
terms of the summation in Eq. (9) have  lower contributions. 
Furthermore,  since $\sum_{q\in \mathbb{Z}} \: q^{-4} $ converges, 
the summation in Eq. (9) converges. 

Figure \ref{fig:PolesBelow} compares the poles below the resonance, for the given example in Fig. \ref{fig:epsallzones}, 
using a numerical technique (Muller's method) and the poles expressions derived earlier in Eqs. (\ref{eq:zone1}, \ref{nearres}). 
Both results show an excellent agreement in zones 1 and 2, while the derived expressions 
can be used  asymptotically to evaluate the poles in between these zones.
\begin{figure}    
  \includegraphics[width=1\linewidth, keepaspectratio]{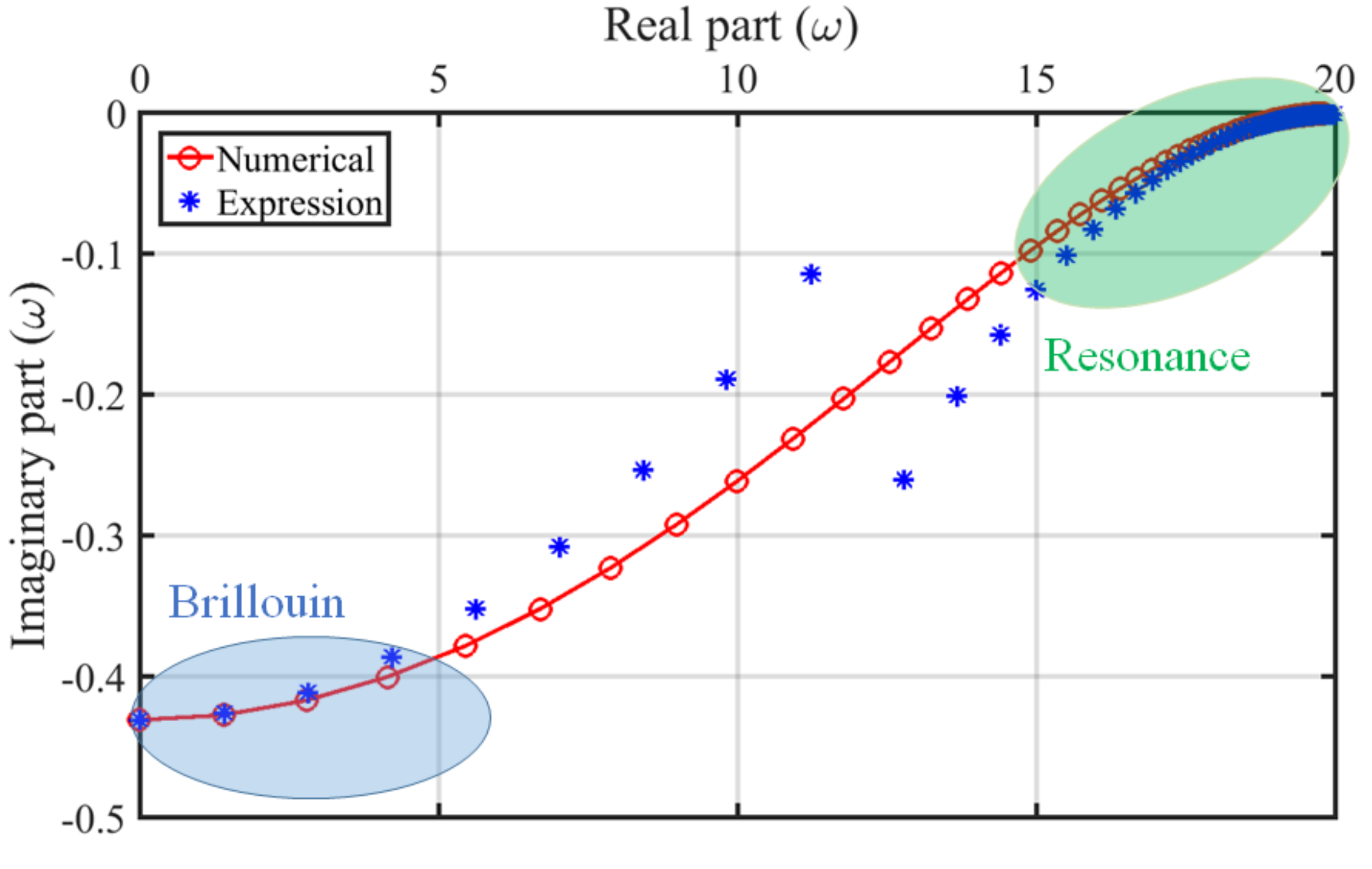}
 \caption[Optional caption]{The poles below resonance for a Lorentz medium of $\om_0 = 20 $ and $\Om=40$. 
 The poles expressions for zone 1 and 2 show excellent agreement with the numerical calculations.}
 \label{fig:PolesBelow}
 \end{figure}

%------------------------------------------------------------------------------------------------
\subsubsection{[Zone-3]: Near Plasma frequency ($\ep \simeq 0$)} \label{sec:AppPlas}
%------------------------------------------------------------------------------------------------
For the near-plasma zone  $\om \simeq \om_p$, the parameter $\zeta^2 \ll 1 $ is introduced so that,
\begin{equation}\label{epsapp}
\omega_{q} = \omega_p ( 1+ \zeta^2) \, . 
\end{equation}  
Then the permittivity becomes
\begin{equation}\label{epsapp}
\ep(\omega)  \simeq \frac{2 \omega_p^2}{\Omega^2} \, \zeta^2  \quad 
\Longrightarrow \sqrt{\ep(\omega)} \approx \frac{\sqrt{2} \omega_p}{\Omega} \, \zeta 
\end{equation} 
and, using that this permittivity is near zero, 
the reflection coefficient is 
\begin{equation}\label{epsapp}
\rho(\om) \approx 1 - \frac{2\sqrt{2} \omega_p}{\Omega} \, \zeta  \, .
\end{equation} 
Taking the logarithm of  Eq. (\ref{eq:MainApp}) leads to
\begin{equation}\label{epsapp}
- \frac{2\sqrt{2} \omega_p}{\Omega} \, \zeta = i q \pi - i \frac{\sqrt{2} \omega_p^2}{\Omega} \, \zeta ( 1+ \zeta^2) 
\end{equation}
and, using the presumption of $\zeta^2 \ll 1$, it is obtained that
 \begin{equation}\label{epsapp}
\zeta = \frac{q \pi \Omega}{\sqrt{2} \omega_p ( \omega_p^2 +4)} (\omega_p  - 2i) \, .
\end{equation}
Since the plasma wavelength of materials are mostly in the 
nanometer region \cite{dressel2002electrodynamics}, then the slab thickness can be assumed to be much larger 
than the plasma wavelength. In this situation, the normalized plasma frequency can be 
safely assumed to be $\omega_p \gg 1 $. 
Finally, we reach an expression for the poles near the plasma frequency $\omega_p$,
 \begin{equation}\label{eq:zone3}
\omega_{q}=\omega_p \; \left[ 1+ \frac{q^2 \pi^2 \Omega^2}{2 \omega_p^4}  \;(1 - 4 i / \omega_p) \right] .
\end{equation}  
The limit of validity for this expression is $ q \ll \frac{\omega_p^2}{ \Omega}$ 
in order to satisfy $\zeta^2 \ll 1$.

\begin{figure}    
  \includegraphics[width=1\linewidth, keepaspectratio]{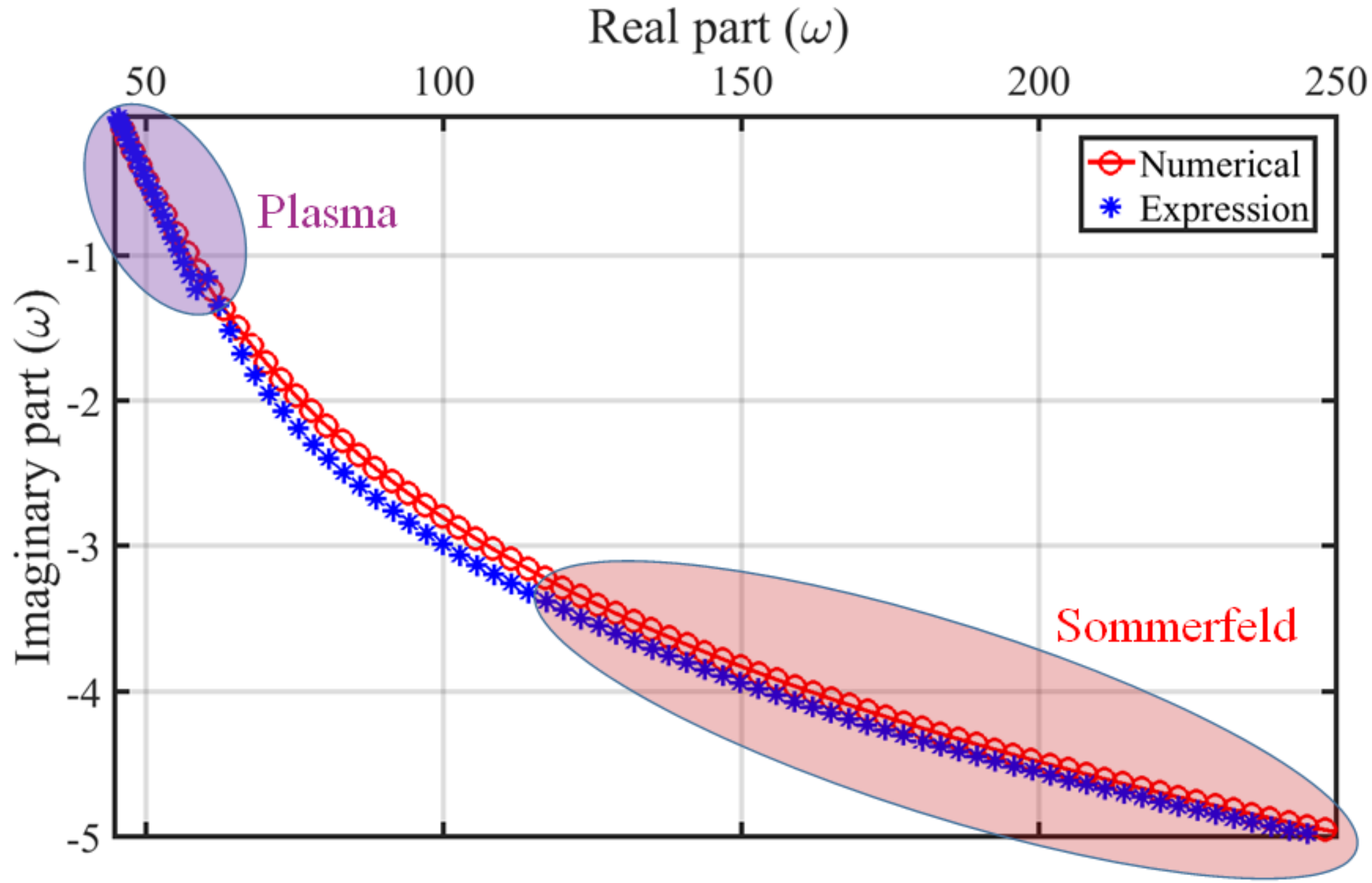}
 \caption[Optional caption]{The poles above plasma frequency  for a Lorentz medium of  
 $\om_0 = 20 $, $\Om=40$ and $ \om_p = 44.7$. 
 The poles expressions for zone 3 and 4 are compared with  numerical calculations.}
 \label{fig:PolesAbove}
 \end{figure}

%------------------------------------------------------------------------------------------------
\subsubsection{[Zone-4] High Frequency (Sommerfeld precursor)} \label{sec:AppFar}
%---------------------------------------------------------------------------------------
The medium cannot follow the excitation of the source when  $\om \gg \omr$   and behaves like the vacuum  with a group velocity $v_g \to c$. 
Thus, the contribution of the poles of this zone appears at the onset of the transmitted field from the dispersive media. \\
The permittivity is close to unity  $\ep \simeq 1$, which leads to  
\begin{equation}\label{sqrteps}
\sqrt{\epsilon(\omega)} \simeq 1 - \frac{\Omega^2 /2 }{\omega^2-\omr^2} \, ,
\end{equation}
and then implies
\begin{equation}\label{rho}
\rho(\omega)  \simeq \frac{\Omega^2 / 4}{\omega^2-\omr^2}  \simeq 0 \, .
\end{equation}
Hence, from (\ref{A:a}) and (\ref{A:b}), it is obtained that
\begin{equation}\label{ab-parts}
a(q) = {q\pi} \, , \quad
b (q)=  \ln{\frac{q^2 \pi^2 -\omr^2}{\Omega^2/4}} \, .
\end{equation}
{\revb{
After a second iteration, the poles expression is
\begin{equation}\label{eq:zone4}
\omega_{q} = {q\pi} + \frac{\Omega^2 /2}{q \pi} - i \ln{\frac{q^2 \pi^2 -\omr^2}{\Omega^2/4}} \, .
\end{equation} 
}}
This expression for frequency poles remains valid as soon as 
$\pi^2 q^2 \gg \omr^2 + \Omega^2 /4 $. 
%%%%%%%%%%%%%%%%%%%%

Figure \ref{fig:PolesAbove} shows the poles above the plasma frequency 
using both the numerical technique and the expressions derived in Eqs. (\ref{eq:zone3}, \ref{eq:zone4}). 
Both results show an excellent agreement in zones 3 and 4.
%------------------------------------------------------------------------------------------------
\renewcommand{\theequation}{B.\arabic{equation}}
\setcounter{equation}{0}

\renewcommand\thefigure{B.\arabic{figure}}    
\setcounter{figure}{0} 

\section{ Effect of absorption\label{absorp}}
%------------------------------------------------------------------------------------------------
\begin{figure} [h]
  \includegraphics[width=1\linewidth, keepaspectratio]{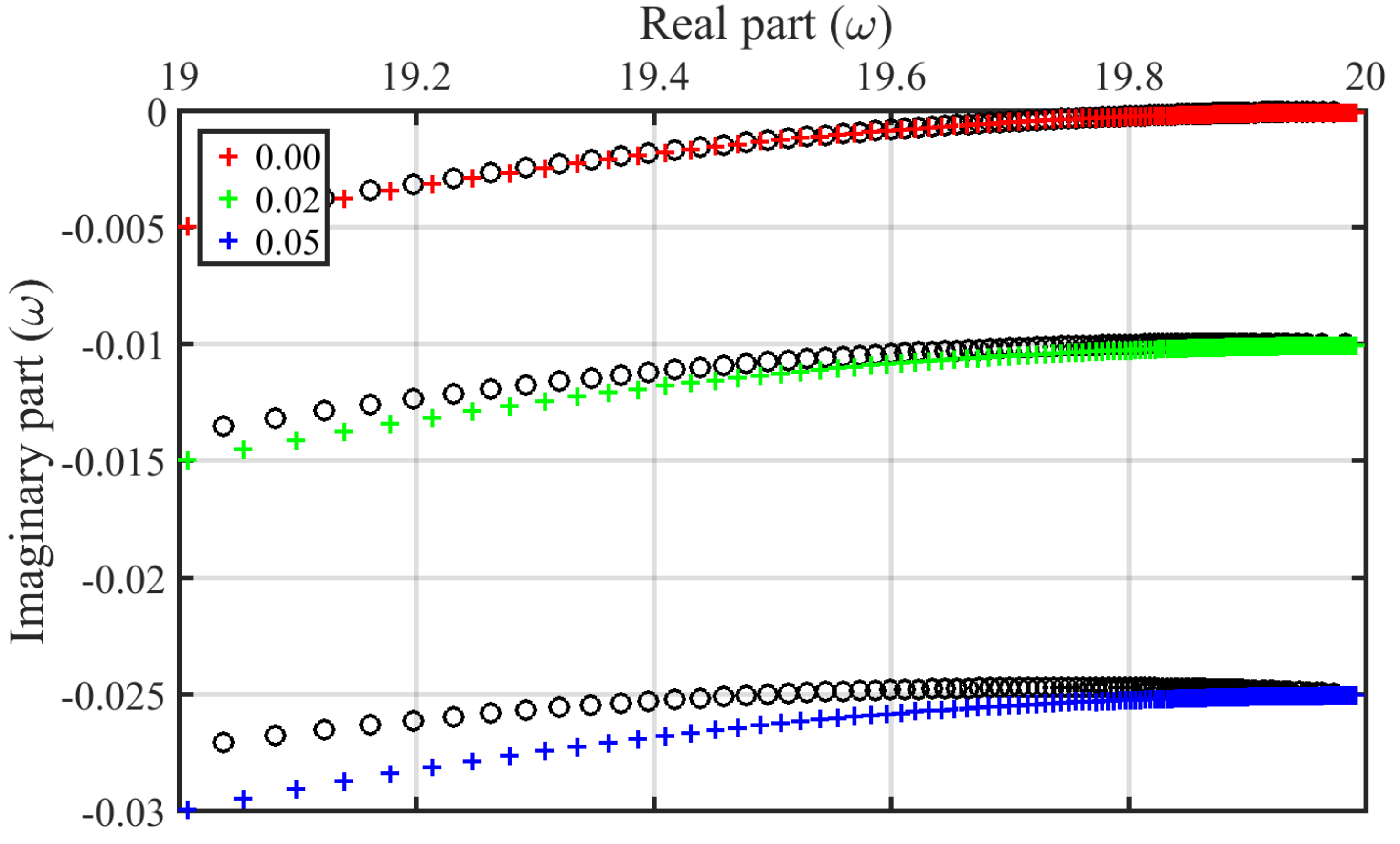}
 \caption[Optional caption]{The poles near the resonance frequency of a Lorentz medium of 
 $\om_0 = 20 $ and $\Om=40$ for different values of the absorption coefficients 
 $\gamma= 0$, $0.02$ and $0.05$. 
 The poles expressions of zone 2 for the lossy case [+, Eq. (\ref{eq:PolesLossy})]
 is compared with  numerical calculations [$\circ$].}
 \label{fig:PolesAbs}
 \end{figure}

The effect of absorption on the poles of a dispersive resonator is briefly discussed. 
The absorption indicates an energy transfer between the incoming wave and the medium. Assuming a weakly-absorbing dispersive medium ($\gamma \ll \omr$), it can be expected to have a significant absorption in the resonance region. In this case, the expression of poles in zone 2 needs to be modified. The absorption leads to a change in the imaginary part of the poles of the resonator. The modified  expression for the poles can be obtained near  resonance  using the lossy expression of the permittivity provided in Eq. (\ref{epDL}), by replacing $\omr$ by $\omr- i \gamma/2$  in the expression of the poles of zone 2: Eq. (\ref{nearres}). The poles are then given by 

%%%%%%%%%%%%%%%%%%%%
\begin{equation} \label{eq:PolesLossy}
\omega_{q} \approx \omr -  \frac{ \omr  \Omega^2 }  { 2 q^2 \pi^2 } \left[ 1+i \frac{4\omr}{q^2\pi^2}+ 
i \gamma \frac{q^2 \pi^2}{\omr \Omega^2} \right].
\end{equation} 
In this case, the imaginary part of the poles is affected by both the medium
absorption and the reflectivity of the interfaces of the slab. 
Figure \ref{fig:PolesAbs} compares the poles near-resonance obtained by numerical 
analysis and by the expression of the lossy case given by Eq. (\ref{eq:PolesLossy}) for different values of the absorption coefficient. Both results are matched in the zone near the resonance.

%\bibliography{references} % bibliography data in report.bib

%

%\bibliographystyle{spiebib} % makes bibtex use spiebib.bst

%\begin{thebibliography}{9}
%\bibliographystyle{ieeetr}
%\end{thebibliography}

%\newpage
%\clearpage

%blablacar
\end{document}